\documentclass[prd,preprint,nofootinbib,superscriptaddress]{revtex4}

\usepackage{graphicx, cancel}
\usepackage{amsmath,amssymb,graphicx,psfrag}

\begin{document}

\newcommand{\gsim}{ \mathop{}_{\textstyle \sim}^{\textstyle >} }
\newcommand{\lsim}{ \mathop{}_{\textstyle \sim}^{\textstyle <} }

%%%%%%%%%%%%%%%%%%%%%%%%%%%%%%%%%

\renewcommand{\thefootnote}{\fnsymbol{footnote}}

\title{Gluon jet as a probe of a long-lived colored particle at the LHC}

\renewcommand{\thefootnote}{\alph{footnote}}

\author{Motoi Endo}
\affiliation{Department of Physics, University of Tokyo, Tokyo 113-0033, Japan}
\author{Shinya Kanemura}
\affiliation{Department of Physics, University of Toyama, Toyama 930-8555, Japan }
\author{Tetsuo Shindou}
\affiliation{Department of General Education, Kogakuin University, Tokyo 163-8677, Japan}
\preprint{UT-HET 029}
\preprint{UT-09-20}
\preprint{KU-PH-002}
\begin{abstract}
\noindent
In some new physics models, there exists a long-lived colored particle.
Although such a particle is expected be discovered by studying the muon-like tracks, 
it is not easy to discriminate hadronic events from leptonic ones at the Large Hadron 
Collider. We focus on the charged track events associated with a single hard gluon jet. 
They are sensitive to the colored long-lived particle and found to be almost free from 
the background after applying the velocity cut. We also study the the process to probe 
properties of the particle.
\end{abstract}
\maketitle
%%%%%%%%%%%%%%%%%%%%%%%%%%%%%%%%%
%\section{Introduction}
%%%%%%%%%%%%%%%%%%%%%%%%%%%%%%%%%
At the Large Hadron Collider (LHC) a huge number of new particles would be produced 
from $pp$ collisions, if physics beyond the Standard Model (SM) exists in the TeV scale. 
In a class of new physics models, a heavy colored particle becomes (quasi) stable and does 
not decay in the detector. In supersymmetric standard models (SSM), the scalar partner of 
the top quark (stop) can be lightest among the supersymmetric (SUSY) particles. Then, the 
stop does not decay into the other SUSY particles as long as the R-parity is conserved. 

The stable stop readily forms so-called R-hadrons\cite{Gates:1999ei,Mackeprang:2009ad}. 
The hadron would be discovered even at the early stage of the LHC. In fact, if it has an electric 
charge and reaches the muon detector, it is observed as a heavy muon-like event. 
Although the light quarks within the R-hadron strongly interact with matter in the detector, 
their momentum fraction is very small. Thus, the energy loss is estimated to be $O(1-10)$ 
GeV for the R-hadron mass $> O(100)$ GeV\cite{Aad:2009wy}, and the R-hadron is likely to 
reach the muon detector. 

The heavy charged-track events are triggered and distinguished from the muon events when 
the velocity of the long-lived particle, $\beta$, is lower than 1. The ATLAS collaborations study 
such events and plan to measure the velocity. The strategy to measure $\beta$ of the R-hadron 
is the same as that of the leptonic track\cite{Aad:2009wy}\footnote{
The velocity is also measured by the CDF collaboration. The absence of the exotic signal 
of the heavy muon up to $\sigma_{\text{obs}}\simeq 48\;\text{fb}$\cite{Aaltonen:2009kea} 
leads to the bound on the stop mass as $m_{\tilde{t}_1}\gtrsim 200\;\text{GeV}$, depending 
on the efficiency factor. }. 
Based on the standard ATLAS reconstruction packages, the velocity will be measured with 
an efficiency $\gsim 0.8$ for $\beta \gsim 0.7$\cite{Aad:2009wy}. On the other hand, another 
method is developed with refining the Resistive Plate Chambers measurement and using the 
Monitored Drift Tubes data\cite{Tarem:2009zz}. This gives a better reconstruction of $\beta$ 
for a lower velocity, providing the efficiency larger than $0.8$ for $\beta \gsim 0.4$ in the case 
of the long-lived R-hadron. The ATLAS packages simultaneously measure the mass of the 
long-lived charged particle\cite{Aad:2009wy,Tarem:2009zz}.

The signatures in the muon detector, however, are not suitable for probing the color property 
of the long-lived particle. It is even difficult to identify whether the particle is colored or not 
model-independently. It has been proposed that we study the track in the inner detector\cite{Aad:2009wy}. Since hadrons strongly interact with the detector matters, they may change its charge 
though passing the calorimeters. Thus, it can happen that the hadron is electrically neutral in 
the inner detector, and then, turns to be charged in the calorimeters. Then, we may be able to 
distinguish the hadronic events by searching for the muon-like ones without tracks in the inner 
detector. However, the interactions of the R-hadron, namely the charge-conversion process, 
strongly depends on the hadron models (see \cite{Mackeprang:2009ad} for a detector simulation 
based on a Regge-based model of the R-hadrons). Therefore, it is difficult to study the color 
properties in a model-independent manner. 

In this Letter, we propose an alternative way to explore the property of the long-lived colored 
particle. We consider the event of the pair charged tracks of the long-lived particle associated 
with a single hard gluon jet\footnote{The similar process has been studied to search for the 
long-lived colored particle, when the hadron is electrically neutral\cite{RhadronStudy}.}. The 
heavy muon-like tracks are triggered by measuring its velocity\cite{Aad:2009wy,Tarem:2009zz}. 
Then, measurements of the associated gluon jet can provide a complementary information to 
the previously proposed analysis\cite{Aad:2009wy}. We will show that the gluon emission is 
effective only when the (quasi) stable particle is colored. Additionally, we will see that the 
momentum distribution of the gluon jet is sensitive to the mass of the long-lived particle. 

\begin{figure}
\begin{center}
\begin{minipage}{5cm}
\includegraphics[scale=0.7]{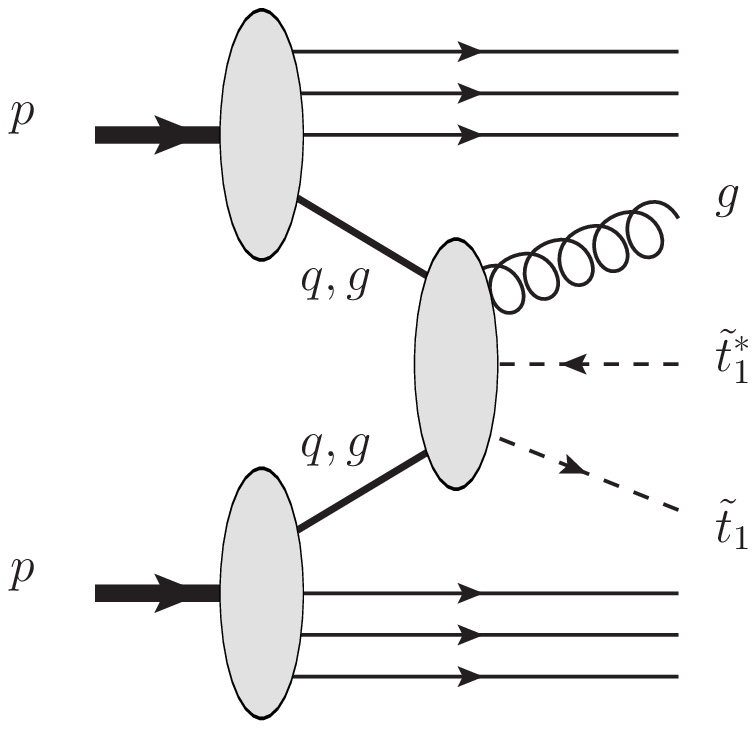}
\end{minipage}
\phantom{Space}
\begin{minipage}{5cm}
\begin{tabular}{c}
\includegraphics[scale=0.6]{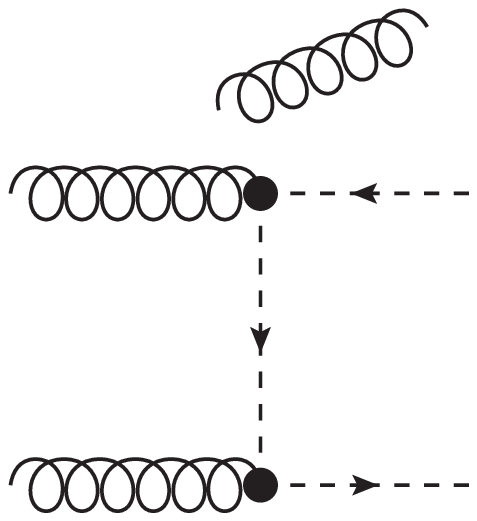}\\
\includegraphics[scale=0.6]{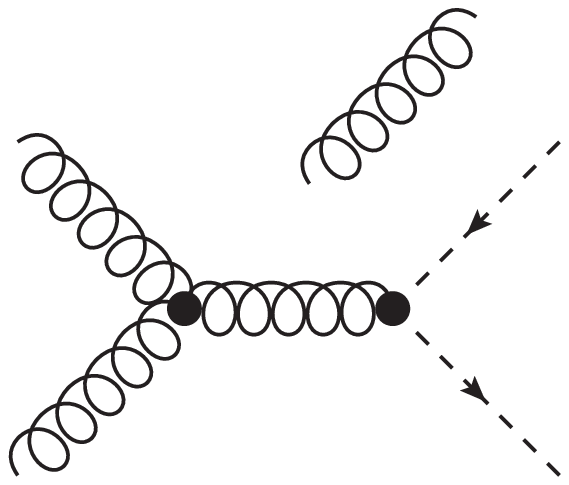}\\
\includegraphics[scale=0.6]{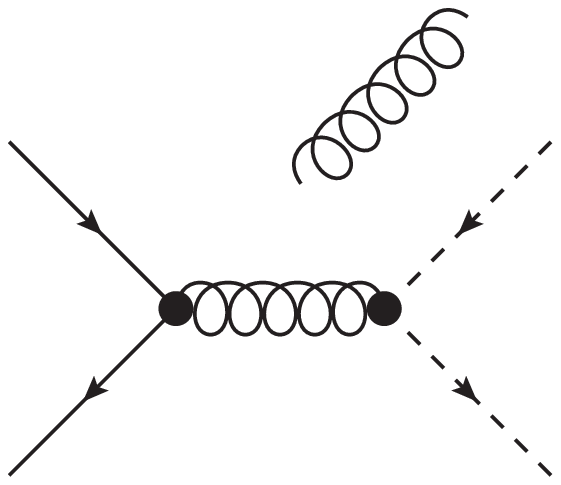}\\
\end{tabular}
\end{minipage}
\end{center}
\caption{The diagrams for the stop pair production associated with a gluon from the $pp$ collision 
(left) and its sub-processes (right).}
\label{fig:feynmandiagram}
\end{figure}

In the following, we study the long-lived stop scenario for definiteness. The stop pair production 
associated with the single jet, $gg(q\bar{q})\to \tilde{t}_1\tilde{t}_1^*g$, is displayed in Fig.~\ref{fig:feynmandiagram}. Then, the observed cross section of the pair production of the long-lived 
stop is expressed as $\sigma_{\text{obs}} = \kappa\lambda \sigma_{\text{prod}}(pp\to \tilde{t}_1
\tilde{t}_1^*g)$, where $\sigma_{\text{prod}}$ is the production cross section by the $pp$ collisions. 
The efficiency factor $\lambda$ denotes a probability that the hadron reaches the muon detector 
with electrically charged, which depends on the hadron models and the property of the colored 
particle, while $\kappa$ represents the others including the detector effects. In the numerical 
analysis, we use {\tt CalcHEP} 2.5.1 \cite{CalcHEP} to calculate $\sigma_{\text{prod}}
(pp\to \tilde{t}_1\tilde{t}_1^*g)$ as well as the SM contributions at the leading order. 

The velocity distribution of the long-lived particles is distinctively different from that of the muon 
events. In fact, the SM background is dominated by the $pp \to \mu\bar\mu jj$ events with 
mis-identifying one of the two jets, e.g. $qg \to qg + A^*/Z^*(\to \mu\bar\mu)$ with $A^*$ and 
$Z^*$ denoting off-shell. They can be suppressed by applying the velocity cut. Actually, if we 
require the jet being hard and away from the beam directions, it is noticed that at least one 
muon is likely to have $\beta \sim 1$ from the kinematical point of view. 

\begin{figure}
\begin{center}
\begin{tabular}{cc}
\includegraphics{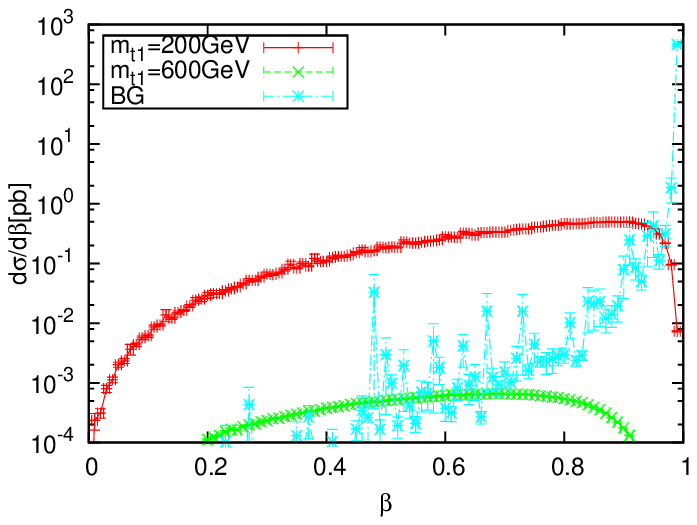}&
\includegraphics{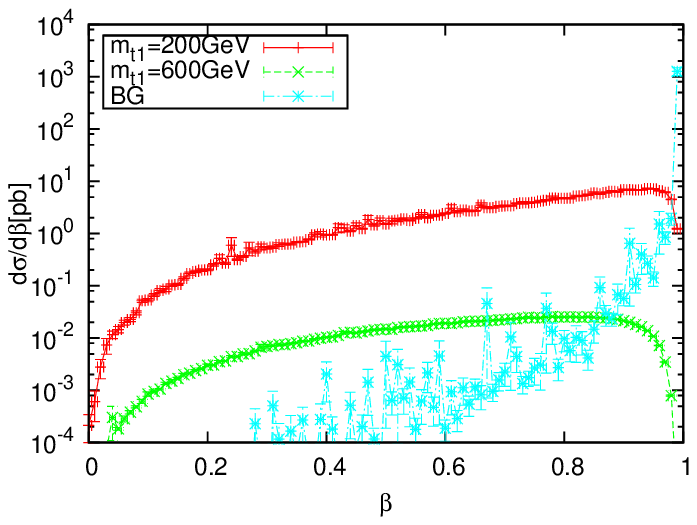}\\
(a)&(b)\\
\end{tabular}
\end{center}
\caption{
The velocity distribution of the differential cross section, $d\sigma(pp\to \tilde{t}_1\tilde{t}_1^*g)
/d\beta$ in the case of $m_{\tilde{t}_1}=200\;\text{GeV}$ and $600\;\text{GeV}$. The signal is 
represented by the solid line, while the dashed line is the SM background. The center-of-energy 
$\sqrt{s}$ is (a) 7 TeV and (b) 14 TeV. Note that we do not impose the velocity cut for either of 
the long-lived particles. If one of the muon has a low velocity, the other of the pair muons tends 
to have $\beta \sim 1$ in the SM background.
}
\label{fig:beta-dist}
\end{figure}

In Fig.~\ref{fig:beta-dist}, we show the velocity distribution of the stops and the muons for the 
signal and the background, respectively. In the figure, we employed the cuts: the angles among 
the tracks, the jet, and the initial beams are in the range of $[10^{\circ}, 170^{\circ}]$, and 
we pick up the events with the gluon energy larger than the cut, $E_{\text{CUT}} = 200$ GeV.
The collision energy is taken to be 7 TeV and 14 TeV. We notice that the stop velocity distributes 
uniformly, while the SM background becomes suppressed for $\beta < 0.8$. Furthermore, 
assuming that one muon has $\beta < 0.8$, the other of the pair muons is most likely to have 
$\beta \sim 1$. Thus, we find that the background becomes almost negligible by cutting the 
velocity of both the long-lived particles. 

As a result, we employ the following cuts in the analysis: 
(i) both the charged-track particles have a velocity $0.4 < \beta < 0.8$, (ii) the angles among 
the tracks, the jet, and the initial beams are in the range of $[10^{\circ}, 170^{\circ}]$, and (iii) 
we pick up the events with the gluon energy larger than the cut, $E_{\text{CUT}} = 200$ GeV. 
Then, the events of the pair charged tracks with the hard gluon jet is background free. 
Note that in addition to the above background, the QCD background is avoided by the angular 
cut (ii), because the muons are produced in a decay of heavy hadrons, and the muons tend 
to distribute around the jet\cite{Aad:2009wy}. Also, the event, $pp \to t\bar t \to\mu\bar\mu\nu
\bar\nu b\bar b$, can be sufficiently suppressed. 

Let us comment on the velocity cut. In the analysis, we applied the lower bound, $\beta > 0.4$, 
which is obtained from Ref.\cite{Tarem:2009zz}. However, the velocity cut $\beta \gsim 0.7$ 
may give a better timing coincidence of the signal of the hard gluon jet in the calorimeter with 
those of the charged tracks in the muon detector. On the other hand, we do not need to measure 
the velocity so precisely to trigger the $gg(q\bar{q})\to \tilde{t}_1\tilde{t}_1^*g$ event on the 
contrary to the methods in the literature\cite{Aad:2009wy,Tarem:2009zz}. This is because the 
following analysis does not rely on details of the charged tracks as long as they are identified 
as the non-muon events. Thus, a full detector simulation is required to estimate the efficiency, 
and we just use the velocity cut given above. 

\begin{figure}
\begin{center}
\includegraphics{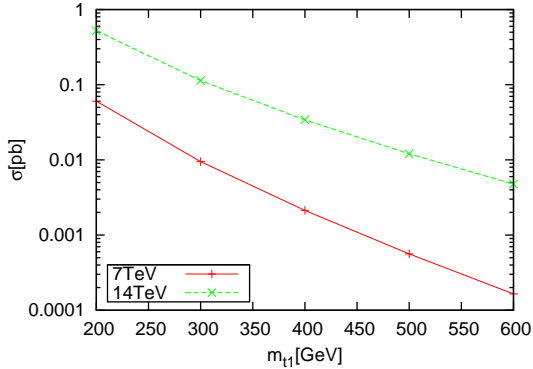}
\end{center}
\caption{
The cross section $d\sigma(pp\to \tilde{t}_1\tilde{t}_1^*g)$ after the cuts for varying the stop mass. 
The SM background is negligibly small. 
}
\label{fig:sigmappt}
\end{figure}

In Fig.~\ref{fig:sigmappt}, we show the production cross section after applying the cuts (i)--(iii). 
The event is almost background free and large enough to observe even at the early stage of 
the LHC. We want to emphasize that the result is characteristic to the colored particle. In fact, 
it can be checked that if we consider a leptonic long-lived particle, the cross section of the 
gluon-associated event is estimated to be three orders of magnitudes smaller than that of the 
colored event. Thus, we can distinguish the colored track from the non-colored one\footnote{
Signals of such a long-lived charged particle at the LHC are studied in the literature\cite{stau}.
}.
Similarly, it is checked that the cross section of the pair muon events with the single jet production 
is negligible under the cuts. 

\begin{figure}
\begin{center}
\begin{tabular}{cc}
\includegraphics{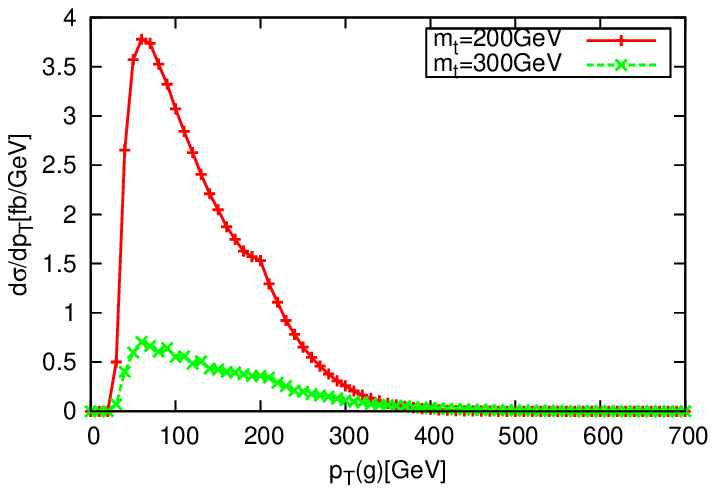}&
\includegraphics{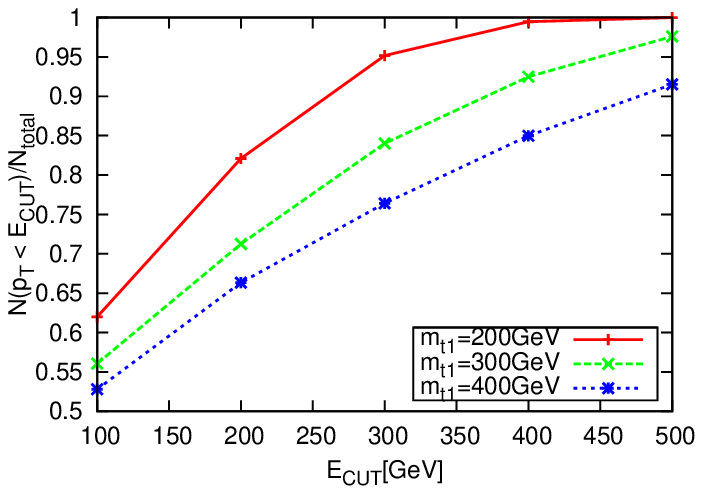}\\
(a)&(b)\\
\end{tabular}
\end{center}
\caption{
(a) the differential cross section $d\sigma(pp\to \tilde{t}_1\tilde{t}_1^*g)/dp_T$ 
as a function of $p_T$ after the gluon energy cut at $E_{\text{CUT}}=200\;\text{GeV}$. 
Here, the solid line is for $m_{\tilde{t}_1}=200\;\text{GeV}$, and the dashed line is for 
$m_{\tilde{t}_1}=300\;\text{GeV}$.
(b) the $m_{\tilde{t}_1}$ and $E_{\text{CUT}}$ dependence of the ratio of the 
number of the event with the gluon transverse momentum $p_T<E_{\text{CUT}}$ 
to the total event number. The solid, dashed, and dotted lines are for 
$m_{\tilde{t}_1}=200\;\text{GeV}$, $m_{\tilde{t}_1}=300\;\text{GeV}$,
and $m_{\tilde{t}_1}=400\;\text{GeV}$, respectively.
The center of mass energy, $\sqrt{s}$, is taken as 14\,TeV.
}
\label{fig:ttG}
\end{figure}

Next, we study the transverse momentum distribution, especially paying attention to the cross 
section in the range of $p_T < E_{\text{CUT}}$. In Fig.~\ref{fig:ttG} (a), we plot the differential 
cross section, $d\sigma/dp_T$. It is noticed that the kink exists around the cut energy. This is 
because, since the energy of the gluon jet is restricted to be $E>E_{\text{CUT}}$, the angler 
distribution of the jet with $p_T < E_{\text{CUT}}$ tends to be limited, while the jet with $p_T > 
E_{\text{CUT}}$ is distributed in a wide range of the angle. 

The cross section below the cut energy is sensitive to the mass of the stop on the contrary to 
that above $E_{\text{CUT}}$. From Fig.~\ref{fig:ttG} (a), we notice that $p(T)(g)$ is likely to 
distribute below $E_{\text{CUT}}=200\,\text{GeV}$ for a lighter stop. In Fig.~\ref{fig:ttG} (b), 
we vary the cut energy $E_{\text{CUT}}$ and show its dependence of the ratio for three different 
stop masses, $m_{\tilde{t}_1}=200\;\text{GeV}$, $300\;\text{GeV}$ and $400\;\text{GeV}$. We find 
that the ratio of the event number with $p_T<E_{\text{CUT}}$ to the total event number, 
$N(p_T<E_{\text{CUT}})/N_{\text{total}}$, is sensitive to $m_{\tilde{t}_1}$. Since the mass 
dependence of the ratio $N(p_T<E_{\text{CUT}})/N_{\text{total}}$ is determined by the kinematics 
of the final states, the QCD corrections does not change the result significantly, although the total 
cross section can be affected.

%\section{Conclusion}

Let us comment on the possible charge transmission of the hadron in the detector, which is related 
to the estimation of the efficiency factor, $\lambda$. It is non-trivial whether the long-lived charged 
hadron reaches the muon detector or not. As we have discussed above, the produced stop charged 
hadrons may change to the neutral one in the calorimeter. Additionally, when a charged hadron is 
heavier than a neutral hadron, a produced charged hadron can decay into the neutral one. However, 
when the mass difference is less than a few MeV as is expected from charm hadrons, the transition 
time is longer than a few seconds, and the charged hadrons can reach the muon 
detector\cite{Gates:1999ei}. Moreover, apart from the discussion above, the charged hadrons may 
be trapped in the hadron calorimeter. This process is rather model dependent, and based on an 
analysis with a Regge model, some sort of the charged hadrons, {\it e.g.} a hadron from the gluino, 
tends to stop in the hadron calorimeter\cite{Mackeprang:2009ad}. Nonetheless, it is claimed that the 
stop charged hadron can reach the muon detector.

In conclusion, a gluon jet is an useful probe of a long-lived colored particle at the LHC even 
at the early stage with low energy/luminosity. A combined study of our analysis with that in 
Ref.~\cite{Aad:2009wy} can determine the color property of the charged track as well as the 
mass of the stop and the hadron property.

%\section*{Acknowledgment}
The work of SK was supported, in part, by Grant-in-Aid for Scientific Research (C),
Japan Society for the Promotion of Science (JSPS),  No. 19540277.
%%%%%%%%%%%%%%%%%%%%%%%%%%%%%%%%%

%%%%%%%%%%%%%%%%%%%%%%%%%%%%%%%%%
\end{document}